\documentclass[aps,prl,twocolumn,floatfix,longbibliography, superscriptaddress]{revtex4-1}
\usepackage{xcolor}

\newcommand*{\colorboxed}{}
\def\colorboxed#1#{%
	\colorboxedAux{#1}%
}
\newcommand*{\colorboxedAux}[3]{%
	\begingroup
	\colorlet{cb@saved}{.}%
	\color#1{#2}%
	\boxed{%
		\color{cb@saved}%
		#3%
	}%
	\endgroup
}

\usepackage{pifont}

\usepackage{float}
\usepackage{epsfig}
\usepackage{bm}
\usepackage[matrix,frame,arrow]{xy}
\usepackage[applemac]{inputenc}
\usepackage[T1]{fontenc}
\usepackage{lmodern}
\usepackage[english]{babel}
\usepackage{amsmath}
\usepackage{ae}
\usepackage{amssymb}
\usepackage{color}
\usepackage{graphicx}
\usepackage{bbm}
\usepackage{url}
\usepackage{nomencl}
\usepackage{subfigure}
\usepackage{slashed}

\usepackage{fixmath}
\usepackage{booktabs}

\newcommand{\ket}[1]{|#1\rangle}

\newcommand{\ketbra}[2]{\left| #1 \rangle \langle #2 \right|}

\newcommand{\nn}{\nonumber}

\renewcommand{\eqref}[1]{\mbox{Eq.~(\ref{#1})}}

\newcommand{\be}{\begin{equation}}
\newcommand{\ee}{\end{equation}}
\newcommand{\bea}{\begin{eqnarray}}
\newcommand{\eea}{\end{eqnarray}}

\usepackage{xr}
\usepackage[colorlinks]{hyperref}
\hypersetup{%
	plainpages=true,
	breaklinks=true,
	hypertexnames=false,
	pageanchor=true,
	colorlinks=true,
	linkcolor={blue},
	citecolor={red},
	urlcolor={blue},
	anchorcolor={black}
}

\newcommand{\beq}{\begin{eqnarray}}
\newcommand{\eeq}{\end{eqnarray}}

\begin{document}
	
	
	\title{Resolution of Gauge Ambiguities in Ultrastrong-Coupling Cavity QED}
	
	
	\author{Omar Di Stefano}
	\affiliation{Theoretical Quantum Physics Laboratory, RIKEN Cluster for Pioneering Research, Wako-shi, Saitama 351-0198, Japan}
	
	\author{Alessio Settineri}
	\affiliation{Dipartimento di Scienze Matematiche e Informatiche, Scienze Fisiche e  Scienze della Terra, Universit\`{a} di Messina, I-98166 Messina, Italy}

	\author{Vincenzo Macr\`{i}}
	\affiliation{Theoretical Quantum Physics Laboratory, RIKEN Cluster for Pioneering Research, Wako-shi, Saitama 351-0198, Japan}
	
	\author{Luigi Garziano}
	\affiliation{Theoretical Quantum Physics Laboratory, RIKEN Cluster for Pioneering Research, Wako-shi, Saitama 351-0198, Japan}
	
	\author{Roberto  Stassi}
	\affiliation{Theoretical Quantum Physics Laboratory, RIKEN Cluster for Pioneering Research, Wako-shi, Saitama 351-0198, Japan}

	\author{Salvatore Savasta}
	\email[corresponding author: ]{ssavasta@unime.it}
	\affiliation{Theoretical Quantum Physics Laboratory, RIKEN Cluster for Pioneering Research, Wako-shi, Saitama 351-0198, Japan}
	\affiliation{Dipartimento di Scienze Matematiche e Informatiche, Scienze Fisiche e  Scienze della Terra,
		Universit\`{a} di Messina, I-98166 Messina, Italy}
	
	\author{Franco Nori}
	\affiliation{Theoretical Quantum Physics Laboratory, RIKEN Cluster for Pioneering Research, Wako-shi, Saitama 351-0198, Japan} \affiliation{Physics Department, The University
		of Michigan, Ann Arbor, Michigan 48109-1040, USA}
	

	\begin{abstract}
		Gauge invariance is the cornerstone of modern quantum field theory \cite{Aitchison1989,Wilczek2005, Maggiore2005,Wilczek2013}.
		Recently, it has been shown that the quantum Rabi model, describing the dipolar coupling between a two-level atom and a quantized electromagnetic field, violates this principle \cite{DeBern2018,DeBernardis2018,Stokes2018}.
		This widely used model describes a plethora of quantum systems and physical processes under different interaction regimes \cite{Kockum2018,Forn-Diaz2018}.
		In the ultrastrong coupling regime, it provides predictions which drastically depend  on the chosen gauge.
		This failure is attributed to the finite-level truncation of the matter system. We show that a careful application of the gauge principle is able to restore gauge invariance even for extreme light-matter interaction regimes. The resulting quantum Rabi Hamiltonian in the Coulomb gauge differs significantly from the standard model and  provides the same physical results obtained by using the dipole gauge. It contains field operators to all orders that cannot be neglected when the coupling strength is high.
		These results shed light on subtleties of gauge invariance in the nonperturbative and extreme interaction regimes, which are now experimentally accessible, and solve all the long-lasting controversies arising from gauge ambiguities in the quantum Rabi and Dicke models \cite{Hepp1973, Wang1973, Rza1975,Lambert2004, Keeling2007,Nataf2010a,Vukics2014,Grie2016,Bamba2016b,DeBern2018}. 
		

	\end{abstract}
	
	\pacs{ 42.50.Pq, 42.50.Ct
	}
	\maketitle
	
	
	
	The ultrastrong-coupling (USC) between an effective two-level system (TLS) and the electromagnetic field has been realized in several solid state systems \cite{Kockum2018,Forn-Diaz2018}. In this regime of quantum	light-matter interaction, going beyond weak and strong coupling, the coupling strength becomes comparable to the transition frequencies in the system.	Recently, circuit QED experiments involving a single LC-oscillator mode 
	coupled to a flux qubit superconducting quantum circuit have realized coupling strengths larger then the system transition frequencies \cite{Yoshihara2017,Yoshihara2017b}. This extreme interaction regime has been denoted deep strong coupling (DSC).
	In these interaction regimes, several properties of coupled light-matter
	systems change drastically, opening the way to a wealth of new intriguing physical effects (see, e.g. \cite{Ashhab2010,Casanova2010,Niemczyk2010,Ridolfo2012,Ridolfo2013,Stassi2013,Cirio2016, DeLiberato2009, DeLiberato2014,Garziano2015,Garziano2016, Garziano2017,DiStefano2017,Kockum2017a,Felicetti2018}) which may offer opportunities for the development of quantum technologies \cite{Nataf2010,Nataf2011,Romero2012,Kyaw2015,Wang2016,Stassi2017,Kockum2017, Armata2017,Stassi2018}.
	
	The form of the electron-photon interaction
	is gauge dependent (see, e.g., \cite{Babiker1983}). However, all physical
	results must be independent of this choice.
	Gauge invariance is a general guiding principle in
	building the theory of fundamental interactions (see, e.g., \cite{Aitchison1989, Maggiore2005}).
	Let us consider, for example, a particle field whose action is invariant under a global phase change [$U(1)$ invariance]. If this phase is allowed to depend on the space-time coordinate $x$, its action is not invariant because the coordinate-dependent phase $\phi(x)$ does not commute with the derivatives (corresponding to the four-momentum) in the action. The symmetry can be restored replacing these derivatives  with covariant derivatives: $D_\mu = (\partial_\mu + i q A_\mu)$, where $q$ is the charge parameter and $A_\mu$ is the gauge potential. If the transformation of the particle field $\psi \to e^{i \phi(x)}$ is associated to a gauge  transformation of the four-vector potential $A_\mu \to A_\mu - \partial_\mu \theta$, the local $U(1)$ symmetry is restored.
	The coupling to the  field, obtained performing the replacement $\partial_\mu \to D_\mu$ in the free Lagrangian, is called the ``minimal coupling''.
	
	It has been shown by several authors \cite{Lamb1952,Lamb1987,Starace1971, Girlanda1981,Ismail-Beigi2001} that approximate models for light-matter interactions
	derived in different gauges may lead to different predictions. It has also been shown that the convergence of the two-photon transition rate depends significantly on the choice of the gauge \cite{Bassani1977}.

	When the light-matter interaction becomes very strong, different gauges can lead to drastically different predictions, leading to controversies \cite{Hepp1973, Wang1973, Rza1975,Lambert2004, Keeling2007,Nataf2010a,Vukics2014,Grie2016,Bamba2016b,DeBern2018}.
	For example, in the case of several TLSs interacting with a single mode of an optical resonator \cite{Dicke}, different gauges may even lead to
	very different predictions, such as the presence or the absence  of a 
	quantum phase transition.
	
	One important conclusion that can be drawn
	from these controversies is that, once the light-matter coupling becomes non-perturbative,  the validity of the two-level
	approximation  for the atomic dipoles depends explicitly
	on the choice of gauge. In particular, it has been shown \cite{DeBernardis2018} that
	in the electric dipole gauge (the multipolar gauge in the dipole approximation) the two-level approximation can be performed as long as the Rabi
	frequency remains much smaller than the energies of the off-resonant higher-energy levels. However, it can dramatically fail
	in the Coulomb gauge, even for systems with an extremely anharmonic spectrum \cite{DeBernardis2018}. This unexpected result is particularly unsatisfactory, since the general procedure to derive the multipolar gauge consists of using the minimal coupling replacement first, and then applying the Coulomb gauge \cite{Babiker1983}.
	
	In a recent work \cite{Stokes2018}, considering arbitrary gauges for the interaction between a TLS and the electromagnetic field, it has been shown that physical predictions in the USC and DSC regimes strongly depend on the  gauge choice. The gauge which provides the more accurate results seems to depend on the detuning between the cavity mode resonance and the atomic transition frequency.
	
	From all of these previous studies, it is clear that approximations in the description of the matter system, as e.g., a finite-level truncation, ruin the gauge invariance of the theory. In 1970, Ref.~\cite{Starace1971} pointed out that gauge ambiguities in the calculation of atomic oscillator strengths, originate from the occurrence of nonlocal potentials determined by the approximation procedures.
	Since a {\em nonlocal} potential in the coordinate representation is an integral operator, it does not commute with the coordinate operator. Indeed, it is easy to show that it can be expressed as a {\em local} momentum-dependent operator $V (\hat {\bf r}, \hat {\bf p})$. This  affects the interaction of light with quantum systems described by approximate Hamiltonians. Specifically, in order to introduce the coupling of the matter system with the electromagnetic field, the minimal replacement rule  $\hat {\bf p} \to \hat {\bf \Pi} = \hat {\bf p} - \hat {\bf A}( {\bf r},t)$ ($\hat {\bf A}$ is the vector potential)  has to be applied not only to the kinetic energy terms, but also to the {\em nonlocal potentials} in the effective Hamiltonian of the particles in the system. 
	By applying such a procedure, the ambiguities in the calculation of approximate matrix elements for electric dipole transitions can be removed \cite{Starace1971}.
	Moreover, taking into account the nonlocality of the approximate potential, two-photon transition rates involving Wannier excitons in semiconductors become gauge invariant \cite{Girlanda1981}.
	Also the microscopic quantum theory of excitonic polaritons is affected by the presence of nonlocal potentials. The application of the standard  minimal coupling replacement leads to a total Hamiltonian which, besides the usual $\hat {\bf A} \cdot \hat {\bf p}$ term, displays an additional diamagnetic term $ e^2 \hat{\bf A}^2/ 2m$, where $e$ and $m$ are the electron charge and mass respectively.
	If a limited number of exciton levels are explicitly included in the model, the long wavelength solution of the polariton dispersion cannot be recovered without introducing {\em ad hoc} the Thomas-Reiche-Kuhn sum rule and truncating the summation consistently \cite{Combescot1991}. This procedure presents some ambiguity and appears to be  rather artificial. However, if the nonlocality of the approximate potential is taken into account and the resulting additional terms are included, up to second order in the vector potential, the correct dispersion relation is automatically recovered \cite{Savasta1995,Savasta1996}.
	In summary, the concept of approximation-induced nonlocal potentials, together with an expansion of the interaction Hamiltonian up to second order in the vector potential, was able to overcome a number of gauge ambiguities.

	Here we investigate if this strategy can work in the maximally truncated Hilbert space provided by a TLS, and  in the nonperturbative regimes of cavity QED. This investigation is relevant not only in order to remove gauge ambiguities in quantum optical systems which are attracting great interest, but also provides a general insight on gauge invariance in extreme interaction regimes. 
	We find that {\em the usual strategy}, which consists of taking into account the nonlocality of the atomic potential, performing the minimal coupling replacement and developing the resulting interaction Hamiltonian up to second order in the vector potential, {\em fails when the coupling strength reaches a significant fraction of the resonance frequencies of the system}. 
	We demonstrate that these gauge ambiguities can be eliminated for arbitrary coupling strengths only by taking into account the approximation-induced nonlocality and keeping the resulting interaction Hamiltonian to all orders in the vector potential  $\hat {\bf A}$. The results presented here solve all the long-lasting controversies arising from gauge ambiguities in the quantum Rabi and Dicke models.

	\section{Results}
	
	\subsection{Nonlocal potentials}
	In order to understand why local potentials become nonlocal when the Hilbert space is truncated, let us consider a one-dimensional potential $\hat V$. In the position basis, it can written as 
	\be
	\hat V = \int dx dx'\, \langle x | \hat V | x' \rangle\, \ketbra{x}{x'}
	\ee
	If the potential is local,  its matrix elements can be written as $\langle x | \hat V | x' \rangle \equiv V(x,x') = W(x) \delta (x-x')$. If ${|n \rangle}$ is a complete orthonormal basis, the matrix elements can be expressed as $V(x,x') = W(x) \delta (x-x')=\sum_{n,n'} W_{n,n'} \psi^*_{n'}(x') \psi_n(x)$, where we defined $\psi_n (x) \equiv \langle x|n \rangle$.  Notice that the Dirac delta function can be reconstructed only by keeping all the infinite vectors of the basis. Hence, any truncation of the complete basis can transform a local potential into a nonlocal one. If only two states are included, e.g., the two lowest energy levels, we obtain
	\be\label{nonlocality}
	V(x,x') = W_{1,0}\left[ \psi_0^*(x') \psi_1(x) + \psi_1^*(x') \psi_0(x) \right]
	\ee
	where, for simplicity, we assumed parity symmetry and real matrix elements.
	It is evident that adding the two terms in Eq.(\ref{nonlocality}), which are products of two smooth wavefunctions it is not possible to reproduce the Dirac-delta function and this will result into a potential with an high degree of spatial nonlocality. 
	It has been shown by several authors \cite{Starace1971,Girlanda1981,Ismail-Beigi2001} that a nonlocal potential can be expressed as a momentum-dependent operator $ V( \hat{\bf r}, \hat {\bf p})$. Indeed, by using the translation operator $\psi(x') = \exp{[i(x'-x)  \hat p ]} \psi (x)$, where $\hat p$ is the momentum operator, we obtain
	
	\be
	\int V(x,x') \psi(x') dx' =V(x, \hat p) \psi(x)\, .
	\ee
	\subsection{The dipole gauge}
	
	We consider a single electric dipole with charge $q$ coupled to a single mode of the electromagnetic field. The field mode is described by a harmonic oscillator with
	bare frequency $\omega_c$ and annihilation and creation operators
	$\hat a$ and $\hat a^\dag$. In one dimension, the dipole can be
	modeled as an effective particle of mass $m$ in a potential $V(x)$. Its unperturbed Hamiltonian is $\hat H_0 = \hat p^2/({2 m}) + \hat V(x)$.
	Applying the dipole approximation, the total Hamiltonian in the dipole gauge is
	\be\label{fulldipole}
	\hat H_D = \hbar \omega_c \hat a^\dag \hat a + \hat H_0 
	+ \frac{q^2 A_0^2 \omega_c}{\hbar} \hat x^2 +i q \omega_c \hat x A_0 (\hat a^\dag - \hat a)\, ,
	\ee
	where $A_0$ is the zero point fluctuation amplitude of the vector potential $\hat A = A_0 (\hat a + \hat a^\dag)$.
	
	If the two lowest-energy levels of the effective quantum particle are well separated from the higher energy levels, as in the case of flux qubits \cite{Yoshihara2017}, and if the detuning $\delta = \omega_{10} - \omega_c$ ($\omega_{10}$ is the transition frequency of the two lowest energy levels) is much smaller than the detunings with respect to other transitions,
	it is possible to truncate the Hilbert space considering only the two lowest-energy levels. In this case, the unperturbed particle Hamiltonian reduces to $\hat {\cal H}_0 = \hbar \omega_{10} \hat \sigma_z /2$, and the total Hamiltonian becomes 
	\be\label{RabiD}
	\hat {\cal H}_D =  \hbar \omega_c \hat a^\dag \hat a + \frac{\hbar \omega_{10}}{2} \hat \sigma_z  + i \hbar g_D (\hat a^\dag - \hat a) \hat \sigma_x \, 
	\ee
	where $g_D = \omega_c A_0 d_{10}/ \hbar$ and  $d_{10} \equiv q \langle 1|\hat x | 0\rangle$ is the dipole matrix element. In Eq.~(\ref{RabiD}) we dropped the constant term ${\cal C} = {d^2_{10} \omega_c A_0^2}/{\hbar}$, which results from the expectation value $\langle i|\hat x^2 | j \rangle$ by using the TLS identity operator $\hat I_2 = |0\rangle \langle 0| + |1\rangle \langle 1|$. This term could be calculated more accurately by including it in the particle potential, before the diagonalization. However, we made the choice of considering the interaction terms only after the Hilbert space truncation. De Bernardis et al.~\cite{DeBernardis2018} have shown that, for an effective particle displaying the two lowest-energy levels well separated from the higher energy levels ($\omega_{21} \gg g_D$), the two level approximation provides  results which are in agreement with those obtained using the full Hamiltonian in Eq.~(\ref{fulldipole}), even for extreme coupling strengths ($g_D \gg \omega_{10}$).

	\subsection{Derivation of the quantum Rabi	model in the Coulomb gauge}
	
	In the Coulomb gauge, the Hamiltonian describing the interaction of an effective quantum particle with the electromagnetic field can be obtained applying the substitution $\hat p \to \hat p - q \hat A$ to the particle Hamiltonian $\hat H_0$. The well-known result is
	\be\label{fullRabi}
	\hat H_C = \hbar \omega_c \hat a^\dag \hat a + \hat H_0 
	- i q \hat p A_0 (\hat a^\dag + \hat a) + \frac{q^2 A_0^2}{2 m} (\hat a^\dag + \hat a)^2\, .
	\ee  
	Projecting $\hat H_C$ in a two-level space, we obtain
	\be\label{RabiC}
	\hat {\cal H}'_C =  \hbar \omega_c \hat a^\dag \hat a + \frac{\hbar \omega_{10}}{2} \hat \sigma_z  +  \hbar g_C \sigma_y (\hat a^\dag + \hat a) 
	+ \frac{q^2 A_0^2}{2 m} (\hat a^\dag + \hat a)^2\, ,
	\ee
	where $g_C = g_D\, \omega_{10}/\omega_c$. 
	\begin{figure}
		\centering
		\includegraphics[width=  0.95 \linewidth]{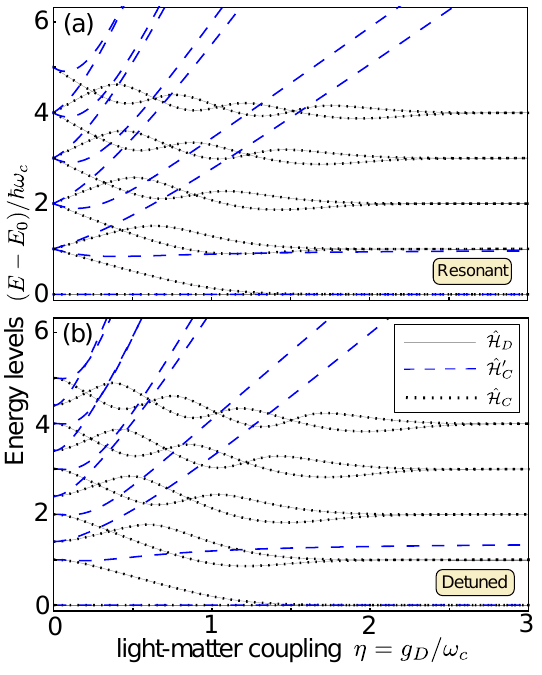}
		\caption{Comparison of the energy
			spectra as a function of the normalized coupling $\eta= g_D /\omega_{c}$, obtained from the quantum Rabi Hamiltonians in the dipole gauge ($\hat {\cal H}_D$), in the Coulomb gauge (standard derivation, $\hat {\cal H}'_C$), and in the Coulomb gauge taking into account the presence of nonlocal potentials ($\hat {\cal H}_C$). The plots in (a) have been obtained using a detuning $\delta=0$; the plots in (b) using $\delta= 0.5 \omega_c$.
			\label{fig:1}}
	\end{figure}
	The diamagnetic term $q^2 \hat A^2/(2m)$ can be absorbed by using a Bogoliubov transformation involving only the photon operators \cite{Yoshihara2017, DeBernardis2018}.
	
	In Fig.~\ref{fig:1} we plot the energy differences ($E_n- E_0$) for the lowest eigenstates
	of $\hat {\cal H}_D$, $\hat {\cal H}'_C$, and $\hat {\cal H}_C$ [see Eq.~(\ref{SSC2})], as a function of the normalized coupling $\eta= g_D /\omega_{c}$. Figure~1a has been obtained at zero detuning ($\delta=0$), while we used $\delta = 0.5\, \omega_c$ for Fig.~1b. The comparison in Fig.~\ref{fig:1} shows that, for very small values of the coupling,  the eigenvalues of the different Hamiltonians reproduce the expected behaviour.
	However, already at moderate coupling strengths, $\eta \sim 0.1$,
	there are significant deviations in the predicted energies, in agreement with the results obtained in  Ref.~\cite{DeBernardis2018}. For values of $\eta \gtrsim 0.5$ the differences become dramatic. The analysis carried out in Ref.~\cite{DeBernardis2018} clarifies that the disagreement between the eigenvalues of $\hat {\cal H}_D$ and $\hat {\cal H}'_C$ is due to the failure of the Rabi model  to provide correct results beyond small coupling values in the Coulomb gauge ( $\hat {\cal H}'_C$).

	However, this derivation of the quantum Rabi model in the Coulomb gauge does not take into account that, in the presence of a truncated Hilbert space, the particle potential can loose its locality $\hat V(\hat x) \to \hat V'(\hat x, \hat p)$. In this case, in order to preserve gauge invariance, one has to apply the substitution $\hat p \to \hat p - q \hat A$ {\em also} to the potential. In principle, this procedure can give rise to additional terms in the interaction Hamiltonian to all orders in the vector potential. These higher order terms are expected to be negligible for small normalized couplings $\eta$, but can become relevant at higher coupling strengths. 
	
	As shown in detail in the Supplementary Note~\textcolor{blue}{1}, using some general operator theorems \cite{Louisell1973}, it is possible to apply the minimal coupling replacement to both the kinetic energy and the nonlocal potential of the effective Hamiltonian of a quantum particle by employing a unitary transformation \cite{Savasta1995}:
	\be\label{SSC}
	\hat H_C = \hat U \hat H_0 \hat U^{\dag} + \hat H_{\rm ph}\, ,
	\ee
	where $\hat H_{\rm ph} = \hbar \omega_c \hat a^\dag \hat a$.
	In the dipole approximation, the unitary operator is
	\be\label{U}
	\hat U = \exp \left[ i \frac{q \hat x \hat A}{\hbar}\right]\, .
	\ee
	
	\begin{figure}
		\centering
		\includegraphics[width=  0.95 \linewidth]{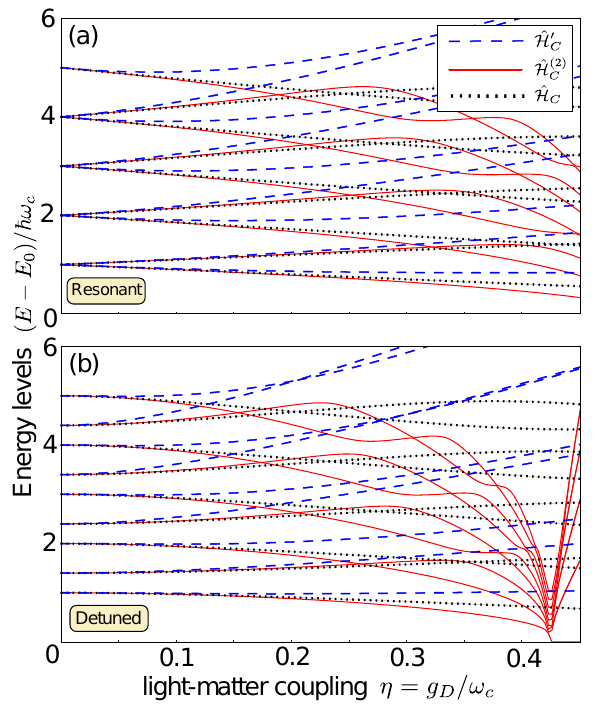}
		\caption{Comparison of the energy
			spectra as a function of the normalized coupling $\eta= g_D /\omega_{c}$, obtained from the quantum Rabi Hamiltonians in the dipole gauge ($\hat {\cal H}_D$), in the Coulomb gauge (standard derivation, $\hat {\cal H}'_C$), and in the Coulomb gauge taking into account the presence of nonlocal potentials ($\hat {\cal H}_C$). The plots in (a) have been obtained using a detuning $\delta=0$; the plots in (b) using $\delta= 0.5 \omega_c$.
			\label{fig:2}}
	\end{figure}
	If the two-level approximation is used, in order to include the interaction, one has to consistently project Eq.~(\ref{SSC}) in the two-level space before performing the unitary transformation of $\hat {\cal H}_0$:
	\be\label{SSC2}
	\colorboxed{red}{
		\hat {\cal H}_C = \hat {\cal U} \hat {\cal H}_0 \hat {\cal U}^{ \dag} + \hat H_{\rm ph}\, ,}
	\ee
	where the projected unitary operator is \be
	\hat {\cal U} = \exp \left[ i \eta \hat \sigma_x (\hat a + \hat a^\dag)\right]\, .
	\ee
	{Note that the standard procedure consists of applying the minimal coupling replacement to the exact matter Hamiltonian $\hat H_0$ as, e.g., in Eq.~(\ref{SSC}), and then to project the resulting Hamiltonian in the truncated Hilbert space: 
		\be\label{Rabl}
		\hat {\cal H}'_C = \hat P  \hat U \hat H_0 \hat U^{\dag}\hat P + \hat H_{\rm ph}
		= \hat P \left[ \hat U \frac{\hat  p^2}{2m} \hat U^{\dag}
		+ \hat V \right] \hat P  + \hat H_{\rm ph}
		\, ,
		\ee
		where 
		$\hat P$ is the projection operator for the truncated Hilbert space, and we used the relation $\hat U \hat V \hat U^\dag = \hat V$, valid when the potential is local. Note that, in contrast to Eq.~(\ref{SSC2}), Eq.~(\ref{Rabl})  contains a nonlocal potential $\hat P \hat V \hat P$ to which the gauge principle has not been applied.
		This analysis explains why the standard approach fails when applied to a matter system described in a truncated Hilbert space.}
	
	Of course, this procedure can be easily generalized to any truncated Hilbert space. It is remarkable that it takes automatically into account the presence of nonlocal potentials, even without knowing them explicitly. 
	Noticing that the operator $\hat{\cal U}$ corresponds to a spin rotation along the $x$ axis, we obtain 
	\be\label{CRabi}
	\colorboxed{red}{\begin{aligned}
			\hat {\cal H}_C &= \hbar \omega_c \hat a^\dag \hat a + \frac{\hbar \omega_{10}}{2}  \\ &\left\{ \hat \sigma_z \cos{\left[ 2 \eta (\hat a + \hat a^\dag)\right]} +
			\hat \sigma_y \sin{\left[ 2 \eta (\hat a + \hat a^\dag)\right]} \right\}
			\, .
	\end{aligned}}
	\ee
	
	This is the correct Hamiltonian in the Coulomb gauge of the quantum Rabi model. The price one has to pay for preserving the gauge principle in such a truncated space is that the resulting Hamiltonian will contain field operators at all orders. The generalization to multimode fields \cite{Male2017,Gely2017,Munoz2018} is straightforward. When the normalized interaction strength is sufficiently weak, it is possible to expand the trigonometric functions up to the linear or quadratic terms. However, when $g_D$ becomes comparable with $\omega_c$ (i.e., the USC regime), such expansion is not sufficient. When approaching, or reaching, the deep strong coupling regime ($\eta \gtrsim 1$) the field terms have to be included at all orders.
	Figure~(\ref{fig:1}) shows that the numerically calculated energy spectrum of $\hat {\cal H}_C$ coincides with that of $\hat {\cal H}_D$. Hence we can conclude that, {\em when the presence of nonlocal potentials, owing to the two-level truncation,  is properly taken into account, it is possible to obtain correct results also in the Coulomb gauge}.
	
	Equation~(\ref{SSC2}) can be easily generalized  beyond the single-mode approximation. Expanding Eq.~(\ref{SSC2}) up to the second order in the potential, we obtain
	\be\label{SSC22}
	\hat {\cal H}^{(2)}_C = \hbar \omega_c \hat a^\dag \hat a 
	+ \frac{\hbar \omega_{10}}{2}  \hat \sigma_z + \hbar g_C \hat \sigma_y (\hat a + \hat a^\dag)+
	\frac{\hbar g^2_C}{\omega_{10}} \hat \sigma_z (\hat a + \hat a^\dag)^2\, .
	\ee
	This equation differs from the standard quantum Rabi Hamiltonian in the Coulomb gauge $\hat {\cal H}'_C$ for the diamagnetic term, which now displays a different coefficient and depends on $\hat \sigma_z$. By introducing  the Thomas-Reiche-Kuhn sum rule in the diamagnetic term of $\hat {\cal H}'_C$ and applying the two-level truncation \cite{Stokes2018}, $\hat {\cal H}'_C$ turns into $\hat {\cal H}^{(2)}_C$. However, this procedure is quite arbitrary. Figure~(\ref{fig:2}) compares the energy spectra of $\hat {\cal H}'_C$,  $\hat {\cal H}^{(2)}_C$, and $\hat {\cal H}_C$. 
	In particular, we observe that for values of the normalized coupling below $0.1$ the spectra of both $\hat {\cal H}^{(2)}_C$ (red lines) and $\hat {\cal H}'_C$  (blue dashed lines) are in quite good agreement with the exact results (expecially those of $\hat {\cal H}^{(2)}_C$) given by $\hat {\cal H}_C$ (black dotted lines). Increasing the coupling, $\hat {\cal H}'_C$ gives spectra which rapidly become very different from the exact energy levels. The spectra of $\hat {\cal H}^{(2)}_C$ provide an improved approximation, which, for the lowest few levels, is acceptable for $\eta \lesssim 0.2$. The range of values of $\eta$ where the approximate models provide acceptable spectra reduces at nonzero detuning [Fig.~(\ref{fig:2}) (b)].
	
	In order to understand how many powers of the photon operators have to be included in  $\hat {\cal H}_C$  to obtain correct spectra at higher $\eta$, we compared in Fig.~3 the approximate spectra, calculated from different $n$-order Taylor expansions $\hat {\cal H}^{(n)}_C$ of   $\hat {\cal H}_C$, with the exact spectra (the eigenvalues of $\hat {\cal H}_C$ ). The results are interesting: for $n=3$ there is already a significant improvement (with respect to $n=2$), up to $\eta \lesssim 0.25$. However, the spectra become completely wrong already at  $\eta \sim 0.3$. Accuracy improves for $n = 10$, but only up to $\eta \lesssim 0.25$. For $n = 200$, there is an excellent agreement, but only for $\eta \lesssim 1.3$. These results show that for values of $\eta$ larger than $1$ (DSC), a very large $n$ is needed to get the correct spectra. However, further increasing $\eta$ requires the inclusion of more and more terms in the expansion. This confirms the {\em nonperturbative spatial nonlocality} which occurs when heavily truncating the particle's Hilbert space.
	
	Specific results for circuit-QED systems can be obtained analogously. For example, the full Hamiltonian of a fluxonium capacitively coupled to an LC oscillator circuit \cite{Manucharyan2009} (corresponding to the charge gauge) can be obtained through an analogous minimal coupling replacement. As shown in the Supplementary Note~\textcolor{blue}{2}, the resulting total Hamiltonian for the two-level model in the charge gauge is very similar to Eq.~(\ref{SSC2}).

	\subsection{Resolution of gauge ambiguities}
	The Hamiltonians $\hat H_D$ and $\hat H_C$ are related by a gauge transformation \cite{Babiker1983}, which can be expressed by the unitary transformation $\hat H_C = \hat {G} \hat H_D \hat {G}^\dag$. In the dipole approximation $\hat {G} = \hat U$ [see Eq.~(\ref{U})].
	Applying this gauge transformation, it is easy to obtain the quantum Rabi Hamiltonian in the Coulomb gauge Eq.~(\ref{fullRabi}) from
	Eq.~(\ref{fulldipole}). Of course, the inverse relationship also holds: $\hat H_D = \hat {G}^\dag \hat H_C \hat {G}$.
	
	As discussed above, it has been shown that $\hat {\cal H}'_C$ (the two-level approximation of $\hat {H}_C$) gives rise to wrong spectra, thus ruining gauge invariance.
	In the previous subsection, we have demonstrated that the correct quantum Rabi Hamiltonian in the Coulomb gauge, based on the complete introduction of the gauge-preserving replacement $\hat {\bf p} \to \hat {\bf \Pi} = \hat {\bf p} - \hat {\bf A}(\hat {\bf r},t)$, is $\hat {\cal H}_C$ in Eq.~(\ref{CRabi}).
	Numerical results in Fig.~\ref{fig:1} show that
	$\hat {\cal H}_C$ gives correct spectra, thus 
	providing clear indications that the procedure developed here restores gauge invariance in TLSs.
	
	We now present an analytical demonstration of the gauge invariance of a TLS coupled to the electromagnetic field. 
	We show that gauge invariance in a truncated Hilbert space can be fully preserved  if the gauge transformation is consistently performed. We start from ${\cal H}_D$, which was shown to be a very good approximation of the full Hamiltonain $H_D$, and apply the gauge transformation projecting the unitary operator $\hat G = \hat U$ in the two-level space:
	$
	\hat {\cal G} \hat {\cal H}_D \hat {\cal G}^{\dag}\, ,
	$ 
	where
	$\hat {\cal G} = \hat {\cal U}$. The result of this unitary transformation should be  $\hat {\cal H}_D \to \hat {\cal H}_C$.
	Noticing that the operator $\hat {\cal U}$ corresponds to a spin rotation along the $x$ axis, and using the Baker-Campbell-Hausdorff lemma, it is easy to obtain
	\be\label{TLgauge}
	\colorboxed{red}{\begin{aligned}
			\hat {\cal G} \hat {\cal H}_D \hat {\cal G}^{\dag} &= \hbar \omega_c \hat a^\dag \hat a  + \frac{\hbar \omega_0}{2}  \left\{ \hat \sigma_z \cos{\left[ 2 \eta (\hat a +  \hat a^\dag)\right]} \right.
			\\ &+ \left.
			\hat \sigma_y \sin{\left[2 \eta (\hat a + \hat a^\dag)\right]} \right\} = \hat {\cal H}_C\, .
	\end{aligned}}
	\ee
	This equation shows that indeed $\hat {\cal G} \hat {\cal H}_D \hat {\cal G}^{\dag} = \hat {\cal H}_C $. This result demonstrates that gauge invariance is preserved in a two-level truncated space, if the gauge transformation is applied consistently, and if we use $\hat {\cal H}_C$ instead of $\hat {\cal H}'_C$. 
	\begin{figure*}
		\centering
		\includegraphics[width= \linewidth]{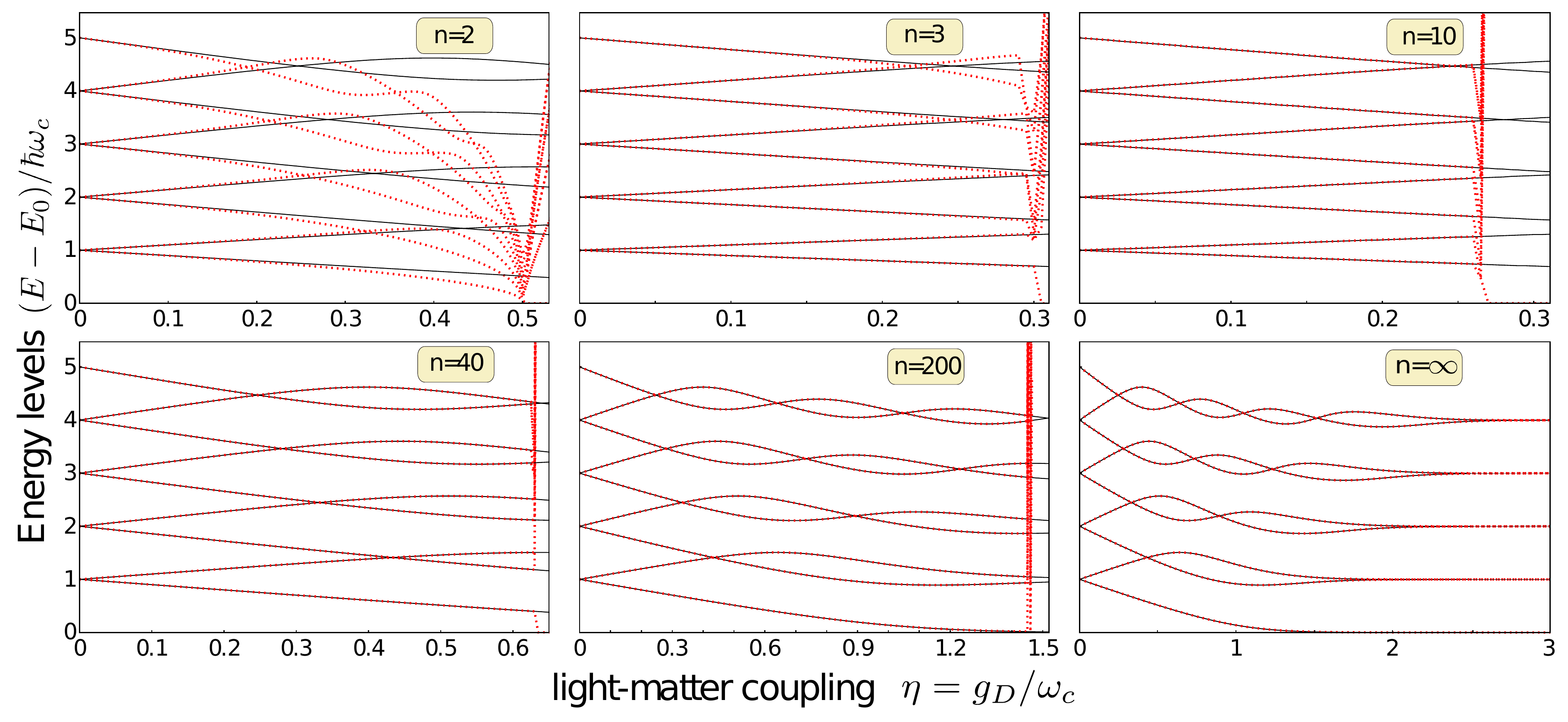}
		\caption{Energy spectra as a function of the normalized coupling $\eta$, obtained from the quantum Rabi Hamiltonians in the Coulomb gauge taking into account the presence of nonlocal potentials ($\hat {\cal H}_C$).
			The panel shows the comparison between the exact (containing all terms) (black continuous curve, see also the bottom-right panel) and approximated energy levels (red curves). The calculations are performed at different order of approximation. The  value $n$ indicates the order of the Taylor expansion used in the approximated Hamiltonian.
			\label{fig:3}}
	\end{figure*}

	Following Ref.~\cite{Stokes2018}, it is possible to employ a formulation in which the gauge freedom is contained within a single real continuous parameter $\alpha$, which determines the gauge through a function $X_\alpha$.
	The general gauge transformation in the dipole approximation is generated by  a unitary transformation determined by the unitary operators
	$
	\hat U_\alpha = \exp \left[ i \hat X_\alpha\right]$,
	where $\hat X_\alpha=\alpha q \hat x \hat A/{\hbar}$.
	The values $\alpha= 0,1$ specify the dipole and the Coulomb gauge, respectively.
	The two-level projected unitary operator is $\hat {\cal U}_\alpha = \exp \left[ i \hat \chi_\alpha\right]$, where $\hat \chi_\alpha=\alpha ({g_D}/{\omega_c}) \hat \sigma_x (\hat a + \hat a^\dag)$.
	The correct $\alpha$-gauge Hamiltonian for a TLS is thus
	\be\label{Ualpha}
	\hat {\cal H}^{(\alpha)} = \hat {\cal U}_\alpha \hat {\cal H}_D \hat {\cal U
	}^{\dag}_\alpha\, .
	\ee 
	We obtain 
	\bea\label{Halpha}
	&&\hat {\cal H}^{(\alpha)} = \hbar \omega_c \hat a^\dag \hat a  -i(1-\alpha) g_D (\hat a - \hat a^\dag)\sigma_x\\ &&+ \frac{\hbar \omega_0}{2}  \left\{ \hat \sigma_z \cos{\left[2 \alpha \eta (\hat a + \hat a^\dag)\right]} + 
	\hat \sigma_y \sin{\left[2 \alpha \eta (\hat a + \hat a^\dag)\right]} \right\} \nonumber .
	\nonumber
	\eea
	Since, the transformation in Eq.~(\ref{Ualpha}) is unitary, the Hamiltonian  (\ref{Halpha}) will have the same energy spectra of $\hat {\cal H}_D = \hat {\cal H}^{(\alpha=0)}$ and of $\hat {\cal H}_C = \hat {\cal H}^{(\alpha=1)}$ for any value of $\alpha$. {\em This eliminates any gauge ambiguity of  the quantum Rabi model}.
	
	\subsection{Derivation of the Dicke model in the Coulomb gauge}
	
	We now extend the results obtained above for a single two-level dipole (Rabi) to the multi-dipole case (Dicke) \cite{Hepp1973, Shammah2017, DeBern2018}. The standard Dicke Hamiltonian in the Coulomb gauge can be written as
	\bea \label{Heff} 
	\begin{split}
		\hat {\cal H}^{\prime N}_{C} & =\hbar \omega_{c} \hat a^{\dag}\hat a + 
		\hbar \omega_{10} \hat J_z\\
		& + 2 \hbar g_c \left(\hat a^{\dag}+\hat a \right) \hat J_y
		+ j\frac{q^{2} A_0^{2}}{m} \left(\hat a^{\dag}+\hat a \right)^{2}\, ,
	\end{split} 
	\eea
	where the number $N$ of dipoles determines the effective main angular momentum quantum number $j = N/2$, and $2 \hat J_i=\sum_{k=1}^N \hat \sigma_{i}^{(k)}$.
	
	Similarly to the quantum Rabi Hamiltonian, it is possible to introduce the coupling between the matter Hamiltonian (in this case ${\cal H}^N_0 = \hbar \omega_{10} \hat J_z$) and the electromagnetic field, by the minimal coupling replacement. In the presence of nonlocal potentials, this can be done by applying the unitary transformation
	\be\label{DickeU}
	\hat {\cal H}^{N}_{C} = \hat {\cal U}_N \hat {\cal H}_0^N {\cal U}^\dag_N
	+ \hat H_{\rm ph}\, ,
	\ee
	where 
	\begin{equation} 
	\hat {\cal U}_N =\exp \! \left [i 2 \eta (\hat a^{\dag}+\hat a) \hat J_x \right ]
	\end{equation} 
	is the projection of 
	$\hat U_N = \exp \! \left[ i {q \sum_k \hat x_k \hat A}/{\hbar}\right]$ in a reduced Hilbert space, which is the tensor product of $N$ two-dimensional Hilbert spaces. We obtain
	
	\be \label{DickeC}
	\colorboxed{red}{\begin{aligned}
			\hat {\cal H}^{N}_{C} = \hbar \omega_{c} \hat a^{\dag}\hat a
			&+ \hbar \omega_{10} \left \{\hat J_z \cos \left[ 4 \eta (\hat a^{\dag}+\hat a)\right] \right. \\&+ \left. \hat J_y \sin \left[ 4 \eta (\hat a^{\dag}+\hat a)\right]\right\}\, . 
	\end{aligned}}
	\ee
	This is the correct Hamiltonian in the Coulomb gauge for the Dicke model. Its gauge invariance can be easily proved following the same procedure shown in the previous subsection.
	{\em This result eliminates any gauge ambiguity} \cite{Keeling2007, Nataf2010a, DeBern2018} {\em of the Dicke model}.
	
	{The Hamiltonian in Eq.~(\ref{DickeC}) is equivalent to the one in Eq.~(45) of Ref.~\cite{DeBern2018} (if direct dipole-dipole interactions are not included). The latter has been obtained applying a polaron transformation to the Dicke Hamiltonian in the dipole gauge. This unitary transformation, as observed by the authors ~\cite{DeBern2018}, is equivalent to a gauge transformation and hence gives rise to a Dicke Hamiltonian in the Coulomb gauge. However, as pointed out there, Ref.~\cite{DeBern2018} does not solve the gauge ambiguities of the Rabi and Dicke models (see also Ref.~\cite{DeBern2018,DeBernardis2018}). Indeed, their direct derivation of the Dicke Hamiltonian in the Coulomb gauge was obtained employing the standard minimal coupling substitution, instead of  the correct generalized one proposed here [Eq.~(\ref{DickeU}]}.

	\section{Discussion}
	
	Approximate models for light-matter interactions derived in different gauges have produced drastically different predictions when the light-matter interaction strength reaches high values, such as the presence or absence of a quantum phase transition. {\em We succeeded to identify  the origin of these gauge ambiguities  recently pointed out}.
	
	Specifically, we found that these ambiguities originate from the presence of  potentials depending on both position and momentum (nonlocal potentials) in the effective Hamltonians  describing the matter systems. This is an unavoidable consequence of the usual approximations (like Hilbert space truncation). In these cases, gauge invariance of the light-matter Hamiltonian can be preserved, even in the presence of extreme light-matter interactions, only by applying the minimal coupling replacement to {\em both} the kinetic and the potential terms of the effective Hamiltonian.
	A careful application of the gauge principle generates a  quantum Rabi Hamiltonian in the Coulomb gauge,  which is significantly different from the standard quantum Rabi model. It contains field operators to all orders that cannot be neglected when the coupling strength is high. The derivation presented here is based on the fundamental minimal coupling replacement, and thus is not limited to the Coulomb gauge. We have also derived 
	arbitrary gauge descriptions (depending on a real parameter $\alpha$)
	of cavity-QED Hamiltonians for TLSs. {\em These results resolve all the controversies arising from gauge ambiguities in the quantum Rabi and Dicke models.}
	
	This analysis bring a deeper understanding of the
	subtleties of  the gauge principle, which cannot be ignored  in the nonperturbative and extreme interaction regimes,  which are now experimentally accessible.  The lesson that can be learned  is very general:  when introducing interactions according to the gauge principle, any approximation in the description of a quantum system requires a {\em generalization  of the minimal coupling replacement} in order to avoid inconsistencies and completely wrong predictions.
	{We presented a general method for the derivation of gauge-invariant light-matter Hamiltonians in truncated Hilbert spaces.}
	This approach can be extended to study ultrastrong and deepstrong light-matter interactions beyond the dipole approximation \cite{Ismail-Beigi2001}, where the multipolar gauge \cite{Babiker1983} is also affected by the presence of nonlocal potentials. A further interesting development  could be the application of this analysis  to  many-body quantum systems strongly interacting with the electromagnetic field \cite{Cortese2017}.
	\vspace{0.4 cm}
	
	\acknowledgments{We acknowledge conversations with S. De Liberato and P. Rabl. FN is supported in part by the: MURI Center for Dynamic Magneto-Optics via the Air Force Office of Scientific Research (AFOSR) (FA9550-14-1-0040), Army Research Office (ARO) (Grant No. 73315PH), Asian Office of Aerospace Research and Development (AOARD) (Grant No. FA2386-18-1-4045), Japan Science and Technology Agency (JST) (the ImPACT program and CREST Grant No. JPMJCR1676), Japan Society for the Promotion of Science (JSPS) (JSPS-RFBR Grant No. 17-52-50023, and JSPS-FWO Grant No. VS.059.18N), RIKEN-AIST Challenge Research Fund, and the John Templeton Foundation.}
	
	The authors declare no competing interests

	


\section*{Supplementary Note 1: Generalized minimal coupling replacement}
In this section, by using some general operator theorems \cite{Louisell1973},  we show how  to implement the  minimal coupling replacement on a generic operator $O(\hat x, \hat p)$ by a unitary transformation \cite{Savasta1995}.
Given two noncommuting operators $\hat \alpha$ and $\hat \beta$ and a parameter $\mu$ we want to calculate $e^{\mu \hat \beta} O(\hat \alpha) e^{-\mu \hat \beta}$.
The function $O(\hat \alpha)$ can be expanded in a power series:
\be\label{expan}
O(\hat \alpha)=\sum_n c_n \hat \alpha^n\, .
\ee
Using \eqref{expan}, we have 
\be
e^{\mu \hat \beta} O(\hat \alpha) e^{-\mu \hat \beta}=
\sum_n c_n e^{\mu \hat \beta} \hat \alpha^n e^{-\mu \hat \beta}\,.
\ee
Observing that 
\bea
e^{\mu \hat \beta} \hat \alpha^n e^{-\mu \hat \beta}=\nn
&&e^{\mu \hat \beta} \hat \alpha e^{-\mu \hat \beta} 
e^{\mu \hat \beta} \hat \alpha e^{-\mu \hat \beta}\\ \nn\ldots\,
&&e^{\mu \hat \beta} \hat \alpha e^{-\mu \hat \beta}=
\left(e^{\mu \hat \beta} \hat \alpha e^{-\mu \hat \beta}\right)^n\, .
\eea

We have:
\be\label{opeq}
e^{\mu \hat \beta} O(\hat \alpha) e^{-\mu \hat \beta}=
\sum_n c_n \left(e^{\mu \hat \beta} \hat \alpha e^{-\mu \hat \beta}\right)^n=
O(e^{\mu \hat \beta} \hat \alpha  e^{-\mu \hat \beta})\,.
\ee
We now apply \eqref{opeq} to $ e^{i  \chi(\hat{x})/\hbar} O(\hat x,\hat p)e^{-i  \chi(\hat{x})/\hbar}$. For the sake of simplicity, we consider here the $\rm 1D$ case. Generalization to $\rm 3D$ is straightforward. We obtain
\be \label{wowp}
e^{i  \hat\chi( x)/\hbar} O(x,\hat p)e^{-i  \hat\chi( x)/\hbar} = O( x,e^{i \hat\chi(x)/\hbar}\hat pe^{-i \chi(x)/\hbar})\, .
\ee
Then, by using the Baker-Campbell-Hausdorff Lemma,  we obtain:
\bea\label{epe}
e^{i \hat \chi(x)/\hbar} \hat pe^{-i  \hat\chi(x)/\hbar}&=& \hat p+ \frac{i}{\hbar} [ \hat\chi( x), \hat p]+ \nn \frac{1}{2}\left(\frac{i}{\hbar}\right)^2[\chi(\hat x),[ \chi(\hat x), \hat p]]+\\ &&\ldots= \hat p- \partial_x \hat\chi(x)\, ,
\eea
where we have used the result $[\hat \chi( x), \hat p]=i\hbar\partial_x  \hat\chi(x)$.
In conclusion, using \eqref{epe}, \eqref{wowp} becomes
\be\label{wow}
e^{i  \hat\chi( x)/\hbar} O( x,\hat p)e^{-i  \hat\chi( x)/\hbar}= O( x,\hat p-\partial_ { x} \hat\chi( x))\,.
\ee
Considering now the special function
\be \label{chi}
\hat\chi( x)=  x    \hat A_0\, ,
\ee
where $\hat A_0 \equiv \hat A(x_0)$ is the field potential calculated at the atom position $x_0$, we obtain
\be \label{chi}
\partial_ { x} \hat \chi( x)= \hat A_0\, .
\ee
If we plug this result into Eq.~(\ref{wow}), we get
\be\label{wow2}
e^{i \hat \chi( x)/\hbar} O( x,\hat p)e^{-i \hat \chi( x)/\hbar}= O( x,\hat p- \hat A_0)\, ,
\ee
demonstrating that the unitary transformation in Eq.~(\ref{wow}) corresponds to the application of the minimal coupling replacement in the dipole approximation.

\section*{Supplementary Note 2: Fluxonium {\em LC} oscillator Hamiltonian in the charge gauge}

Here we derive the total Hamiltonian describing a 
fluxonium qubit (TLS) interacting with a superconducting {\em LC} oscillator \cite{Stokes2018, Manucharyan2009}. 
Considering for simplicity the case of a zero-flux offset, the fluxonium Hamiltonian is:
\be\label{hfluxon}
\hat H_{\rm flux}= 4\tilde E_C \hat N^2 +\frac{\tilde E_L}{2}\hat\phi^2-E_J \cos{\hat\phi} \, .
\ee
where $\hat \phi=\hat\Phi/\hat\Phi_0=2e/\hbar\,\hat\Phi $ is the reduced flux operator with conjugate momentum $\hat N=\hat Q/2e$ (the reduced charge), such that $[\hat\phi, \hat N] = i$. In Eq.~(\ref{hfluxon}), $\tilde E_C$, $\tilde E_L$, and $E_J$ are,  respectively, the capacitive,  inductive, and Josephson energies.

The {\em LC} oscillator is characterized by a capacitance $C$ and an inductance $L$. Its  resonance frequency is $ \omega_{c}=1/\sqrt{L C}$ and  its characteristic impedance is $Z=\sqrt{L/C}$. 
The Hamiltonian of the {\em LC} oscillator is
%
\be\label{hlc}
\hat H_{\rm osc}=
\frac{\hat Q^2}{2C}+\frac{\hat \Phi^2}{2L}\, ,
\ee
where $\hat Q$ and $\hat \Phi$ are the charge and the flux operator, respectively.

We now introduce the  reduced flux operator $\hat \varphi= 2 \pi \Phi /\Phi_0$ and the reduced charge operator $\hat \chi = \hat Q/ (2e)$, whose commutator is $[\hat \varphi, \hat \chi]=i$. These can be expanded in terms of the creation and annihilation
operators:
$$\hat \varphi=\varphi_{0}(\hat a +\hat a^\dag)\, ,$$
$$\hat \chi=-i \chi_{0}  (\hat a -\hat a^\dag)\, ,$$ 
where  $\varphi_{0} =\sqrt{2e^2{L}\hbar\Omega}=\sqrt{2\hbar e^2Z}$, 
and $\chi_{0} =\sqrt{{C}\hbar\Omega/(8e^2)}=\sqrt{\hbar /8e^2Z}$.

The Hamiltonian of the LC oscillator can be expressed as
\be\label{hlc}
\hat H_{\rm osc}=4E_{C}\, \hat\zeta^2+ E_{L}\frac{\hat\varphi^2}{2}\, ,
\ee
where $E_{C} = e^2/(2 C)$ and $E_{L} = (\phi_0/2 \pi)^2 /L$.

The capacitive coupling (charge gauge) between the fluxonium and the $LC$ oscillator
can be described by making the substitution
$\hat N \to \hat N+\hat \chi$ in the fluxonium Hamiltonian in \eqref{hfluxon}. The total Hamiltonian then becomes:
\be\label{HC}
\hat H_{C}=4\tilde E_C (\hat N+\hat \chi)^2 -E_J\cos( \hat \phi) +\frac{\tilde E_L}{2}\hat \phi^2+ 4E_{C} \hat\chi^2+ E_L\frac{\hat\varphi^2}{2}\, .
\ee
This replacement is different from the standard {\em minimal coupling replacement}, it  involves two conjugate momenta, instead of a conjugate momentum and a field coordinate. This minimal coupling replacement can also be obtained by applying a unitary transformation to the fluxonium Hamiltonian $\hat H_{\rm flux}$:
\be 
\hat H_{C} = \hat H_{\rm osc} + \hat {R}\, \hat { H}_{\rm flux} \hat { R}^{\dag}\, ,
\ee
where $\hat { R}=\exp[{i\hat\phi\hat\chi}]$. This can be proved by following the procedure described in the next section.


Diagonalizing $\hat H_{\rm flux}$ and then projecting Eq.~(\ref{HC})  in a two-level space, spanned by the eigenstates $\ket{0}$ (ground state) and $\ket{1}$ (excited state), we obtain
\be\label{Hpcharge}
\hat {\cal H}'_{C}= \hat {\cal H}_{\rm flux} + \hbar \omega_c \hat a^\dag \hat a +i\hbar g_C \hat \sigma_y(\hat a -\hat a^\dag)  -
4 \tilde E_C \chi_{0}^2
(\hat a -\hat a^\dag)^2\, ,
\ee
where in the two-level approximation the reduced flux operator becomes
$\hat\phi= \Phi_{10}\hat \sigma_x$ ($\Phi_{10}\equiv \langle e| \hat \phi |g \rangle$ is assumed real), and  $g_C=\omega_{10}\phi_{10}\chi_{0}$. We also have:
\be
\hat {\cal H}_{\rm flux}= \frac{\hbar\omega_{10}}{2} \hat\sigma_z \,,
\ee
with $\hat\sigma_z=\ketbra{1}{1}-\ketbra{0}{0}$.

It can be shown that the total Hamiltonian in Eq.~(\ref{Hpcharge}) is not gauge invariant because its derivation does not take into account the presence of an effective nonlocal potential which originates from the two-level truncation. In analogy with the procedure employed in the main text, {\em the correct gauge invariant total Hamiltonian} can be obtained applying the following unitary transformation to $\hat {\cal H}_{\rm flux}$:
\be\label{mcrS}
\hat {\cal H}_{C} =  \hbar \omega_{c}\, \hat a^\dag \hat a + \hat {\cal R}\, \hat {\cal H}_{\rm flux} \hat {\cal R}^{\dag} \, ,
\ee
where
\be
\hat {\cal R}= \exp\left[{\frac{g_C}{\omega_{10}}\hat\sigma_x(\hat a -\hat a^\dag) }\right]
\ee
is the unitary operator resulting from the truncation in the two-level space of the unitary operator $\hat R$. The minimal coupling replacement in the truncated space \eqref{mcrS} gives

\bea\label{CoulombS}
\colorboxed{red}{\begin{aligned}
		\hat {\cal H}_C &=\hbar \omega_c \hat a^{\dag} \hat a 
		+\frac{\hbar\omega_{10}}{2}
		\left\{\cosh\left[{\frac{2g_C}{\omega_{10}}}(\hat a -\hat a^\dag)\right]\hat\sigma_z+ \right. \\&+ \left.i\sinh\left[{\frac{2g_C}{\omega_{10}}}(\hat a -\hat a^\dag)\hat\sigma_y\right]
		\right\} \nonumber\,. 
\end{aligned}}
\eea
This equation describes the correct (gauge invariant) total Hamiltonian for a fluxonium qubit interacting with an LC circuit in the charge gauge.

\bibliography{RefME}

\end{document}